\documentclass[floatfix,prb,twocolumn,superscriptaddress, showpacs,amsmath,amssymb,footinbib,aps,10pt]{revtex4-1}

\bibpunct{[}{]}{,}{n}{}{}

\usepackage{natbib}
\usepackage[english]{babel}
\usepackage{letltxmacro}
\LetLtxMacro{\ORIGselectlanguage}{\selectlanguage}
\makeatletter
\DeclareRobustCommand{\selectlanguage}[1]{%
  \@ifundefined{alias@\string#1}
    {\ORIGselectlanguage{#1}}
    {\begingroup\edef\x{\endgroup
       \noexpand\ORIGselectlanguage{\@nameuse{alias@#1}}}\x}%
}
\newcommand{\definelanguagealias}[2]{%
  \@namedef{alias@#1}{#2}%
}
\makeatother
\definelanguagealias{en}{english}
\definelanguagealias{English}{english}

\usepackage{graphicx}
\usepackage[usenames,dvipsnames]{color}

\usepackage{bbm}
\usepackage[color=red]{changes}
\usepackage[colorlinks=true,allcolors=blue]{hyperref}
\definechangesauthor[color=red]{LP}
\colorlet{Changes@Color}{red}

\newcommand{\Ham}{\hat{H}}
\newcommand{\tonde}[1]{\left( #1 \right)}
\newcommand{\quadre}[1]{\left[ #1 \right]}
\newcommand{\graffe}[1]{\left\{ #1 \right\}}
\newcommand{\opc}[1]{{\hat{c}^{\protect\phantom \dagger}}_{#1}}
\newcommand{\opcdag}[1]{{\hat{c}^{\dagger}}_{#1}}
\newcommand{\ket}[1]{\left|#1\right\rangle}
\newcommand{\bra}[1]{\left\langle #1\right|}
\newcommand{\opn}[1]{{\hat{n}^{\phantom \dagger}}_{#1}}

\usepackage{layouts}

\DeclareMathOperator{\sign}{sign}

\begin{document}

\title{Backscattering off a driven Rashba impurity at the helical edge}

\author{Lorenzo Privitera}
\affiliation{Institute for Theoretical Physics,  University of W\"urzburg, 97074 W\"urzburg, Germany}

\author{Niccol\`{o} Traverso Ziani}
\affiliation{Dipartimento  di  Fisica,  Universit\`{a}  di  Genova, and SPIN-CNR, Via Dodecaneso 33,  16146  Genova,  Italy}

\author{In\`{e}s Safi}
\affiliation{Laboratoire de Physique des Solides-CNRS-UMR5802. University Paris-Saclay, B\^{a}t.510,91405Orsay, France}

\author{Bj\"orn Trauzettel}
\affiliation{Institute for Theoretical Physics,  University of W\"urzburg, 97074 W\"urzburg, Germany}

\date{\today}

\begin{abstract}
The spin degree of freedom is crucial for both understanding and exploiting the particular properties of the edges of two-dimensional topological insulators. In the absence of superconductivity and magnetism, Rashba coupling is the most relevant single particle perturbation in this system. Since Rashba coupling does not break time reversal symmetry, its influence on transport properties is only visible if processes that do not conserve the single particle energy are included. Paradigmatic examples of such processes are electron-electron interactions and time dependent external drivings. We analyze the effects of a periodically driven Rashba impurity at the helical edge, in the presence of electron-electron interactions. Interactions are treated by means of bosonization and the backscattering current is computed perturbatively up to second order in the impurity strength. We show that the backscattering current is non-monotonic in the driving frequency. This property is a fingerprint of the Rashba impurity, being absent in the case of a magnetic impurity in the helical liquid. Moreover, the non-monotonic behaviour allows us to directly link the backscattering current to the Luttinger parameter $K$, encoding the strength of electron-electron interactions.
\end{abstract}

\maketitle

\section{Introduction}
Since their theoretical prediction~\cite{Kane_PRL05_2,Kane_PRL05_1,bernevig2006quantum} and subsequent experimental discovery~\cite{konig2007quantum}, Quantum Spin Hall (QSH) systems are attracting significant attention in view of their possible applications in spintronics~\cite{Nadj-Perge_Science14,Roth_Science09,Michetti_APL13,Linder_NPhys15,Breunig_PRL18} and topological quantum computation~\cite{Mong_PRX14}. Accordingly, the number of host materials for QSH   systems is  increasing. Additionally to semiconducting heterostructures, such as HgTe/CdTe-~\cite{konig2007quantum} and InAs/GaSb-~\cite{Knez_PRL11} based systems, bismuthene on SiC~\cite{Reis_Science17}, WTe2~\cite{Wu_Science18} and Jacutingaite~\cite{Marrazzo_PRL18} have been shown to be two-dimensional topological insulators. The bulk band gap, and hence the temperature range in which the QSH effect can be seen, is, in the latter materials, substantially enhanced. However, in order to employ the QSH effect in applications, it is crucial to be able to generate one-dimensional helical channels with long coherence length and to develop technologies enabling their manipulation. With this respect, HgTe-based heterostructures still present advantages. In such systems, in fact, the coherence length can reach several micrometers~\cite{Bendias_NL18,Lunczer_PRL19}. Moreover,  the QSH effect can sustain strong magnetic fields~\cite{Scharf_PRB12,Zhang_PRB14,Zhang_SREP15,Ma_Ncomm15} and the possibility to induce superconductivity has been demonstrated~\cite{Bocquillon_Nnano17,Deacon_PRX17}. Recently, quantum constrictions between edges at different sides of the sample have also been realized~\cite{Strunz_Nphys2020}. The combination of these particularities make certain functionalities viable~\cite{Klinovaja_PRB14_1,Klinovaja_PRB14_2,Li_PRL16,Fleckenstein_PRL19}.

The mechanisms allowing for the manipulation of helical edge states share the common trait of enabling the coupling among particles (electrons or holes) with different spin. This necessity essentially emerges due to spin-momentum locking. Several possibilities along these lines have been considered. Magnetic barriers~\cite{Qi_Nphys08,Fu_PRB09,Timm_PRB12,Dolcetto_PSS13,Fleckenstein_PRB16}, that could in principle allow for the implementation of many interesting proposals, have not been experimentally achieved in QSH systems yet. Alternatively, the possibility of locally manipulating Rashba coupling by means of external gates has been analyzed~\cite{Dolcini_PRB11,Dolcini_PRB17}. It opens the way to interesting electron quantum optics applications~\cite{Dolcini_EPJST18,Ronetti_arxiv19}.

At the same time, Rashba coupling at the helical edge has been the object of intense research with the aim of understanding the possible mechanisms that generate backscattering, even in the absence of explicit time reversal symmetry breaking~\cite{Crepin_PRB12,Geissler_PRB14,Xie_PRL16,Kharitonov_PRB17}. While the combination of strictly local spin-orbit coupling, interactions invariant under spin rotations, and unbounded linear spectrum, does not lead to backscattering~\cite{Xie_PRL16}, relaxing any of these conditions can indeed imply a backscattering current~\cite{Kharitonov_PRB17}. This effect manifests itself in two different configurations: a Rashba scatterer in an otherwise standard helical Luttinger liquid~\cite{Crepin_PRB12,Geissler_PRB14,Kharitonov_PRB17} and a normal scatterer in the so called generic interacting helical liquid~\cite{Schmidt_PRL12}. A generic helical liquid is a helical liquid in which the spin quantization axis defined by spin-momentum locking depends on the quasimomentum~\cite{Schmidt_PRL12,Orth_PRB13,Zhang_PRL14,Orth_PRB15}. In both cases  the combined effect of a spin dependent term and an interaction term mixing the single particle states matters.

The possibility of mixing single particle states at the helical edge is not exclusive to electron-electron interactions. Time dependent processes can result in similar effects. Recently, this principle has been used to demonstrate that the effect of current noise in a generic helical liquid in the presence of spin-independent impurities can indeed result in a backscattering current~\cite{Vayrynen_PRL18}.

In this article, we take a complementary point of view. We consider a standard helical liquid in the presence of a periodically driven Rashba scatterer. Moreover, since interactions at the helical edge are material dependent~\cite{Strunz_Nphys2020,Li_PRL15,Vayrynen_PRB16,Stuhler_NPhys19} and potentially important~\cite{Wu_PRL06,Xu_PRB06}, we include them by means of bosonization~\cite{Haldane_PRL81}. We perturbatively compute the backscattering current as a function of the frequency of the driving. We find that, even in the weak interaction regime, the dependence is non-monotonic. This is in striking contrast with the behavior of usual time dependent impurities in single-channel Luttinger liquids~\cite{Feldman_PRB03,Makogon_PRB06,Makogon_PRB07,Safi_PRB19} (we refer to this case as ``magnetic impurity''~\cite{Note_magnetic} throughout the article). Furthermore, we use such non-monotonicity to define a quantity that only depends on the Luttinger parameter $K$, encoding the strength of electron-electron interactions. This dependence can in principle allow for a determination of $K$, a notoriously difficult task, without the need of considering power law dependencies.

The rest of the article is structured as follows. In  Sec.~\ref{sec:model}, we present the model and the formalism we use for the computation of the backscattering current. In Sec.~\ref{sec:current}, we present the general result. In Sec.~\ref{sec:kink}, we focus on the discussion of the non-monotonic behavior and its implications. We then present the analysis of some relevant limiting cases in Sec.~\ref{sec:limits}. Finally, in Sec.~\ref{sec:conclusion}, we draw our conclusions. Some details of the calculation are described in three appendices.

\section{Model}\label{sec:model}
\subsection{Fermionic representation}The edge states of a QSH insulator are characterized by spin-momentum locking, meaning that right and left movers are uniquely associated to a specific spin polarization~\cite{Wu_PRL06}. We define the fermionic operators $\psi_{R}(x)\equiv \psi_{R\uparrow}(x)$ and $\psi_{L}(x)\equiv\psi_{L\downarrow}(x)$. The edge Hamiltonian in presence of an external bias $V$ can be written as $\Ham_{edge} = \Ham_0 + \Ham_{int} + \Ham_V$ with
\begin{align}
&\Ham_0  = \int \mathrm{d}x \left[-i v_F  \tonde{\psi^\dagger_{R} \partial_x \psi_{R} - \psi^\dagger_{L} \partial_x \psi_{L} }\right] \; ,  \\
&\Ham_{int}  =2\pi v_F g_2 \int \mathrm{d}x \left[ \psi^\dagger_{R}(x)\psi^\dagger_{L}(x) \psi_{L}(x)\psi_{R} (x)\right] \; ,\\
&\Ham_{V}  =  \int \mathrm{d}x  \left[\mu_L\psi^\dagger_{L}(x)\psi_{L}(x) - \mu_R\psi^\dagger_{R}(x)\psi_{R}(x)\right]  \label{eq:H_V}\; .
\end{align}
The helical liquid is characterized by the first two terms: the non-interacting part $\Ham_0$ describes linear dispersing helical fermions with Fermi velocity $v_F$; $\Ham_{int}$ is the only relevant  contact-interaction term allowed by spin-momentum locking if the system is not at half-filling~\cite{Wu_PRL06}.
The external bias enters our description through $\Ham_V$, with $\mathrm{e}V = \mu_L - \mu_R$~\cite{Feldman_PRB03}, with $\mathrm{e}$  the electron charge; for clarity, we take $\mu_R=0$. We imagine that before the addition of any other perturbation, the system has reached a quasi-equilibrium state with respect to the external leads. These leads  are assumed to be  space-separated enough that it makes sense to assign a different chemical potential to each of the fermionic species, mimicking the effect of a finite voltage~\cite{Pecca_PRB03}.  Additionally, we introduce the perturbation,  a single time-dependent Rashba impurity. The associated Hamiltonian is~\cite{Geissler_PRB14} $\Ham_R=\int dx \mathcal{H}_R(x)$, with
\begin{equation}\label{eq:Rashba_Ham1}
\mathcal{H}_R(x) = \alpha(x,t)  \quadre{\partial_x \psi^\dagger_{R}(x) \psi_{L}(x) - \psi^\dagger_{R}(x) \partial_x\psi_{L}(x)+ \rm h.c.} \;,
\end{equation}
where $\alpha(x,t)$, assumed to be real, is the Rashba matrix element.
We consider the impurity to be turned on only after the system has reached quasi-equilibrium with the leads. Moreover, it must be situated far away from both leads. In this case, the quasi-equilibrium distribution functions of left and right movers are not substantially affected, if the driving period is smaller than the electron injection time~\cite{Landauer_RMP94,Fevrier2018}. We assume zero temperature for simplicity.
\subsection{Bosonic representation}
In order to treat  electron-electron interaction in a compact way, we employ bosonization~\cite{Haldane_PRL81,VonDelft_AOP95,Miranda_BJP03}. Fermionic operators are then expressed as
\begin{equation}\label{eq:bosonization}
\psi_{R/L}(x) =\frac{ \kappa_{R/L}}{\sqrt{2 \pi a}} e^{\pm i \sqrt{4 \pi} \Phi_{R/L}(x)} e^{\pm i k_Fx}\;.
\end{equation}
In Eq.~\eqref{eq:bosonization}, $\Phi_{R/L}(x)$ are Hermitian bosonic fields, $\kappa_{R/L}$ are the Klein factors, $k_F$ is the Fermi momentum and $a$ is a short distance cutoff. The usual dual fields are defined as $	\Phi(x) = \, \Phi_{R}(x) + \Phi_{L}(x)$ and $\Theta(x) =  \, \Phi_{R}(x) - \Phi_{L}(x)$ and satsisfy canonical commutation relations $\quadre{\Phi(x), \partial_y\Theta(y)}=-i\delta(x-y)$. The  density of right movers is related to the bosonic fields by $\hat{\rho}_{R} (x) =  \frac{1}{\sqrt{\pi}}\partial_x \Phi_{R}(x)$ and accordingly for left movers. We obtain $\Ham_{LL} = \Ham_0 + \Ham_{int}$ in the standard form
\begin{equation}\label{eq:Luttinger:Ham}
\begin{split}
\Ham_{LL} = \hbar \frac{v}{2}\int \mathrm{d}x\! \quadre{\frac{1}{K}\tonde{\partial_x \Phi(x)}^2 + K \tonde{\partial_x \Theta(x)}^2}
\end{split} 	\;.
\end{equation}
The parameter $K=\sqrt{(1-g_2)/(1+g_2)}$ encodes the interaction strength: $0<K<1$ means repulsive interaction, with $K \to 1$ being the non-interacting limit. The Hamiltonian in Eq.~\eqref{eq:Luttinger:Ham}, supplemented by the chirality-dependent chemical potential given in Eq.~\eqref{eq:H_V}, can also be interpreted in terms of the inhomogeneous Luttinger liquid model~\cite{Safi_PRB95,Ponomarenko_PRB95,Maslov_PRB95,Safi_EPJB99,Dolcini_PRB05}, in the case in which the interacting helical liquid is coupled to non-interacting electron reservoirs. The  inhomogeneous Luttinger liquid model describes one dimensional fermions subject to a space-dependent interaction strength which is assumed to be slowly varying on the scale of the Fermi {wavelength}. Such variation\label{key} is then modelled by position-dependent parameters $K$ and $v$.
In helical liquids, the bosonic sound velocity is not renormalized by interactions in the usual Luttinger liquid fashion, i.e. $vK \neq v_F$. This is due to broken Galilean invariance at low energies in a quantum spin Hall system at the edge, related to the linearity of the spectrum. 
In fact, the dependence of   the velocity on the parameter $K$ can be derived as~\cite{Miranda_BJP03,Geissler_PRB15}
\begin{equation}\label{eq:velocity}
	v  = v_F \tonde{1-  \tonde{\frac{1-K^2}{1+K^2}}^2}^{1/2}    \equiv v_F \,\lambda(K)  \;.
\end{equation}
The Rashba Hamiltonian  instead becomes~\cite{Strom_PRL10,Budich_PRL12,Crepin_PRB12,Geissler_PRB15}
\begin{equation}\label{eq:Rashba_Ham3}
\begin{split}
\Ham_R = &  \int \! \mathrm{d}x \;\left[ \alpha(x,t)  i  \frac{\kappa_{L}\kappa_{R}}{\sqrt{\pi}a} \sum_{m=\pm} e^{2imk_Fx}  \right.\\
& \left.\left(  \partial_x \Theta(x)
	e^{i m \sqrt{4\pi} \Phi(x)}
	+ m \frac{e^{i m\sqrt{4\pi} \Phi(x)}}{\sqrt{\pi}a}\right)\right] \; .
\end{split}
\end{equation}
Moreover, the bias term can be written as $\Ham_{V} =\int\! \mathrm{d}x \,\frac{\mu_L}{2\sqrt{\pi}}\tonde{\partial_x \Phi(x) - \partial_x \Theta(x)}$. It can be shown that the presence of this term is equivalent to the following shift of bosonic fields~\cite{Pecca_PRB03,Geissler_PRB15} %
\begin{equation}\label{eq:shift}
\begin{cases}
\Phi \to \Phi + \frac{\omega_0 t }{\sqrt{4 \pi}} \;,\\
\partial_x \Theta \to \partial_x \Theta - \frac{eV}{\sqrt{4\pi}v_F}
\end{cases}
\end{equation}
with $\omega_0 = \mathrm{e}V/\hbar$. We remark that, following the approach of Ref.~\cite{Pecca_PRB03}, we consider the electron reservoirs to be spatially extended and non-interacting. Consequently, we consider the zero-modes of the system to be non-interacting as well, given the influence of the leads~\cite{Geissler_PRB15}. For this reason, $v_F$, rather than $vK$, appears in Eq.~\eqref{eq:shift}.   

\subsection{Form of the backscattering current}
In the absence of impurities, the current passing through a helical liquid is $\mathrm{e}^2V/h$, meaning that the system shows perfect conductance quantization~\cite{Safi_PRB95,Ponomarenko_PRB95,Maslov_PRB95}. Here, we want to consider the impact of a periodically driven Rashba impurity to lowest order in the impurity strength. For simplicity, we  consider a perfectly localized impurity, parametrized by  $\alpha(x,t)  = \alpha_0  \sin(\omega t)\delta(x-x_0)$. The variation of the current is associated with the rate of variation in the number of right and left movers. Therefore, we need to calculate the expectation value of the operator
\begin{align}\label{eq:current_def}
\hat{I}_{BS}(t) \equiv \mathrm{e}\tonde{ \dot{\hat{n}}_{R} - \dot{\hat{n}}_{L}} = 2 \mathrm{e}\dot{\hat{n}}_{R} = -2 \frac{i}{\hbar} \mathrm{e}\quadre{\hat{n}_{R},\Ham_R}   ,
\end{align}
where the (normal ordered) number operator is defined as $\hat{n}_{R} = \int\! \mathrm{d}x \hat{\rho}_{R} (x) = \int \! \mathrm{d}x :\psi^\dagger_{R}(x)\psi^{\phantom{\dagger}}_{R}(x) :\, $.
As we show in Appendix~\ref{app:currop}, the bosonized version of the backscattering currents reads
\begin{equation}\label{eq:curr_op_bos}
\begin{split}
\hat{I}_{BS}(t) &= -\!\int\! \! \mathrm{d}x \; \alpha(x,t)  2 \mathrm{e}   \frac{\kappa_{L }\kappa_{R }}{\sqrt{\pi}\hbar\,a} \sum_{m=\pm}  e^{2imk_Fx}  \\
&  \tonde{  m\,\partial_x \Theta(x)
	e^{i m \sqrt{4\pi} \Phi(x)}
	+ \frac{e^{im \sqrt{4\pi} \Phi(x)}}{\sqrt{\pi}a}} \;.
\end{split}
\end{equation}
We remark that the  shifts of Eq.~\eqref{eq:shift} have to be implemented in the current operator as well.  In order to calculate its expectation value at a generic time $t$, we treat the impurity as a time-dependent perturbation and use a Kubo-like approach, which has been used in similar impurity problems, see for example  Ref.~\cite{Feldman_PRB03,Makogon_PRB06,Vayrynen_PRL18,Safi_PRB19},
\begin{equation}\label{eq:curr_pert}
\langle \hat{I}_{BS}(t)\rangle =  \frac{i}{\hbar} \int_{-\infty}^{t} \mathrm{d}t' \bra{0} \quadre{\Ham_R(t'),\hat{I}_{BS}(t)}  \ket{0} \;.
\end{equation}
In Eq.~\eqref{eq:curr_pert}, $\ket{0}$ is the ground state of $\Ham_{LL}$ in Eq.~\eqref{eq:Luttinger:Ham}, which generates also time-evolution for the operators. If we did not perform any shift in the bosonic fields, $\ket{0}$ should be the ground state of $\Ham_{LL} + \Ham_{V}$.
\section{Analytic result of the backscattering current}\label{sec:current}
We focus on the dc-component of the current, e.g. $\langle \hat{I}^{\mathrm{dc}}_{BS}\rangle = \frac{1}{\tau}\int_{t_1}^{t_1 +\tau} \mathrm{d}t \langle \hat{I}_{BS}(t) \rangle$, where $t_1$ is a generic time after the impurity has been completely switched-on and $\tau$ the driving period. 
As a result of a tedious but straightforward calculation (reported in Appendix~\ref{app:bosonization}) we obtain
\begin{widetext}
\begin{equation}\label{eq:Kubo_4}
\begin{split}
	\langle \hat{I}^{dc}_{BS}\rangle = {}& \mathrm{e}\frac{\alpha_0^2\,}{2\pi \hbar^2 v^{2(K+1) }a^{2(1-K)} }\frac{2K+1}{ K\,\Gamma(2+2K)} \sum_{r=\pm}r \left|\omega + r\omega_0\right|^{2K+1}\sign\left(\omega +r\omega_0\right) + \\
	{}&- \mathrm{e}\frac{\alpha_0^2 }{\pi \hbar^2 v^{2K+1}} \frac{\omega_0}{v_F}\frac{1}{a^{2(1-K)}}\frac{1}{\Gamma(2K+1)} \sum_{r=\pm} \left|\omega +r\omega_0\right|^{2K}  +	 \\
	{}& + \mathrm{e}\frac{\alpha_0^2 }{2\pi \hbar^2 v^{2K}} \frac{\omega_0^2}{2v_F^2} \frac{1}{a^{2(1-K)}}\frac{1}{\Gamma(2K)} \sum_{r=\pm} r\left|\omega +r\omega_0\right|^{2K-1}\sign\left(\omega +r\omega_0\right) \;.
\end{split}
\end{equation}
\end{widetext}
In Eq.~\eqref{eq:Kubo_4}, $t_a = a/v$ is a short-time cutoff and $\Gamma(x)$ is the Euler Gamma function. For the sake of simplicity, we neglected exponential factors of the kind $\exp-(t_a \left|\omega \pm \omega_0\right|)$, because we are considering $\omega, \omega_0 \ll E_g$, with $E_g$ being the bulk gap $E_g$. In our model, the bulk gap $E_g$ can indeed be identified with $\hbar v/a$ (the high-energy cutoff).  It is convenient to introduce the dimensionless parameter	
\begin{equation}
\gamma(K) \equiv \frac{\alpha^2_{\mathrm{ad}}}{4 \pi  K^2\,\Gamma(2K) \,\lambda^{2(K+1)}(K) (a k_F)^{2\tonde{1-K}}} \;,
\end{equation}
which contains a dimensionless version of the impurity matrix element $\alpha
_{\mathrm{ad}} = \alpha_0 k_F/\hbar v_F$, while $\lambda(K)$ is defined in Eq.~\eqref{eq:velocity}. In this way, the backscattering current can be written as
\begin{equation}\label{eq:Kubo_5}
\begin{split}
\langle \hat{I}^{dc}_{BS}\rangle =  \mathrm{e}\tonde{\frac{\hbar}{E_F}}^{2K} \gamma(K) \mathfrak{I}\tonde{K, \omega, \omega_0} \; ,
\end{split}
\end{equation}
where $E_F = \hbar v_F k_F$. The function $\mathfrak{I}$ in Eq.\eqref{eq:Kubo_5}, which has the dimensions $1/s^{2(K+1)}$, is instead given by the sum of three contributions, each one containing different $K$-dependent powers of $\omega\pm\omega_0$:
\begin{equation}
\mathfrak{I} \tonde{K, \omega, \omega_0} = f_{2K+1} + f_{2K} + f_{2K-1}\,
\end{equation}
with
\begin{widetext}
\begin{subequations}\label{eq:Ik}
	\begin{align}
	&f_{2K+1}(\omega, \omega_0) = \sum_{r=\pm}r \left|\omega + r\omega_0\right|^{2K+1}\sign\left(\omega +r\omega_0\right) \;, \\	
	&f_{2K} (\omega, \omega_0) =  -2\lambda K \omega_0 \sum_{r=\pm} \left|\omega + r\omega_0\right|^{2K}  \; ,\  \label{eq:2K}\\	
	\begin{split}
	&f_{2K-1}(\omega, \omega_0) = \tonde{\lambda K\omega_0}^2 \label{eq:f2k-1} \sum_{r=\pm}r \left|\omega + r\omega_0\right|^{2K-1}\sign\left(\omega +r\omega_0\right)  \;.
	\end{split}
	\end{align}
\end{subequations}
\end{widetext}
The analytic expression for the backscattering current represents the main result of the article. It is hence important to comment on its validity. As in the case of the driven magnetic impurity~\cite{Feldman_PRB03}, the divergence of the backscattering current for $K\leq 1/2$, at $\omega=\omega_0$ makes our perturbative result, in the proximity of that point and for strong interactions, unreliable. Moreover, $\langle \hat{I}^{dc}_{BS}\rangle$ should be a small correction to $\mathrm{e}^2 V/h$. This is true for a certain range of $\omega$. As $\omega$ grows, at some point, we leave the validity regime of the perturbation theory.  
\section{Non-monotonicity of $\langle \hat{I}^{dc}_{BS}\rangle$}\label{sec:kink}
	\begin{figure}[t]
			\centering
		\includegraphics[width=0.48\textwidth]{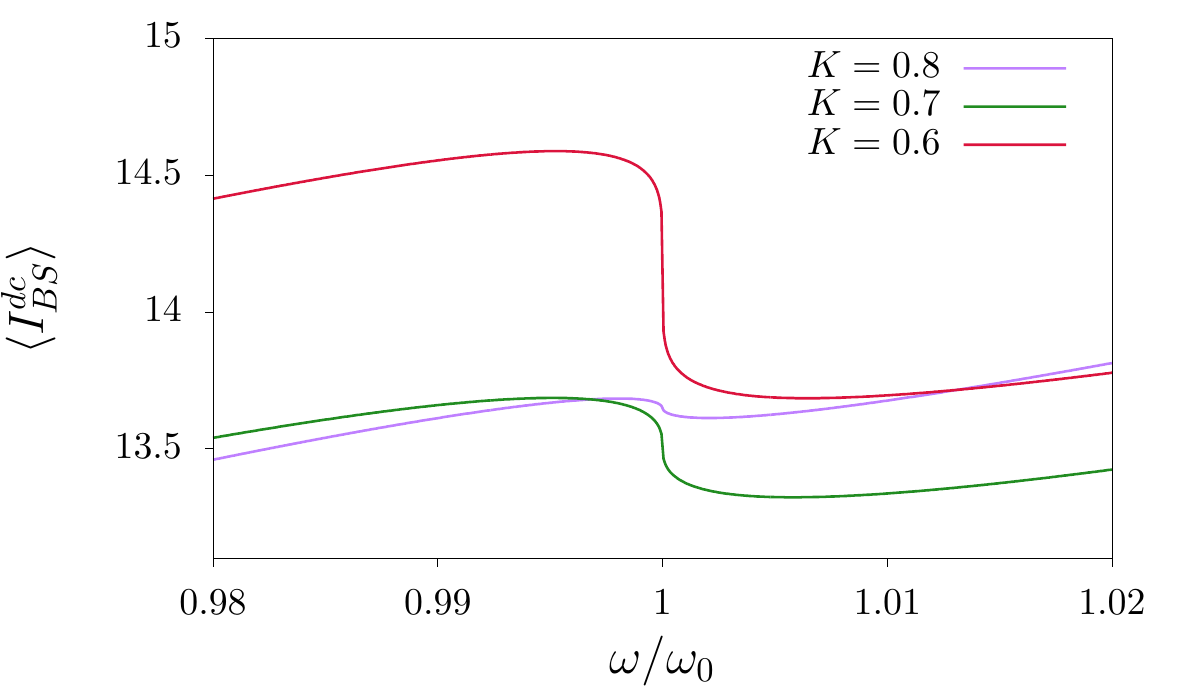}
		\caption{  Zoom around $\omega =\omega_0$ of $\langle \hat{I}^{dc}_{BS}\rangle $, measured in units of $[\mathrm{e}\alpha^2_0k_F^4 \left(k_F a\right)^{(2K-2)}/\left(\hbar^2 \omega_0\right)]$  for $K = 0.8$ (violet), $K=0.7$ (green) and $K=0.6$ (red).  We further set $\hbar = 1$ and   $E_F = 0.1 \hbar\omega_0 $. }
		\label{fig:kink}
	\end{figure}

\begin{figure}[!htp]
	\centering
	\includegraphics[width=0.4\textwidth,angle=0]{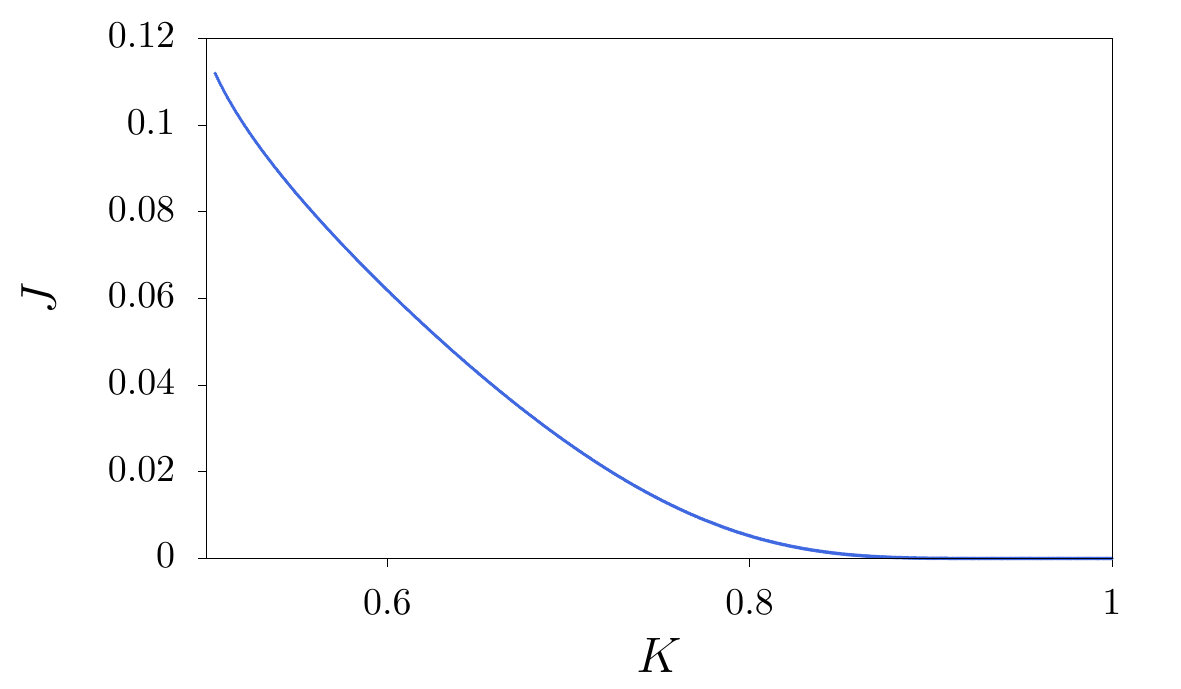}
	\caption{The relative jump defined in the main text as a function of the interaction parameter $K$.}  
	\label{fig:Jump}
\end{figure}
For a moderately interacting helical liquid with $K>1/2$, $\langle \hat{I}^{dc}_{BS}\rangle$ is a continuous function of $\omega$: $f_{2K-1}$ presents a divergence only in its derivatives.  Far away from  $\omega= \omega_0$, $f_{2K-1}$ is negligible and $\langle \hat{I}^{dc}_{BS}\rangle$, as a function of $\omega$, always has a positive derivative. Close to $\omega_0$ this is not  true, because the term proportional to $\left|\omega-\omega_0\right|^{2K-1}$ in $f_{2K-1}$ contributes with a negative diverging derivative. Therefore, $\langle \hat{I}^{dc}_{BS}\rangle$ is not  monotonously growing, but rather exhibits a kink across $\omega_0$, as shown in Fig.~\ref{fig:kink}. To quantify this kink, we call $\omega_{max}$ the local maximum on the left of $\omega_0$,  $\omega_{min}$ the local minimum on the right of $\omega_0$, and  define the relative jump as
\begin{equation}
J \equiv \frac{\langle \hat{I}^{dc}_{BS}\rangle (\omega_{max}) -  \langle \hat{I}^{dc}_{BS}\rangle (\omega_{min})}{\langle \hat{I}^{dc}_{BS}\rangle (\omega_{max})} \;.
\end{equation}
The dependence of $J$ on $K$ is shown in Fig.~\ref{fig:Jump}.  This definition is  convenient, since  all the prefactors to $\mathfrak{I} $ in Eq.~(\ref{eq:Kubo_5}), containing system-specific parameters such as the Fermi velocity, cancel out. Moreover,   $J$ does not depend on the external bias $\mathrm{e}V$, since we can write $\mathfrak{I} (K, \omega, \omega_0) =  \omega_0^{2K+1}\mathfrak{I}'(K, \omega/ \omega_0)  $. Therefore $J$  depends only on the interaction strength $K$.  Thus $J$ can serve as an alternative way to extract $K$ from transport measurements without the need to measure a power-law behaviour. We remark that in the case of a magnetic impurity these considerations do not apply, as the related backscattering current decreases monotonously as $\omega$ increases~\cite{Feldman_PRB03}.
\section{Limits}\label{sec:limits}
\subsection{Static limit}
As a first sanity check, we calculate the static impurity  limit,  $\omega\to 0$, of the backscattering current in Eq.~\eqref{eq:Kubo_4} and compare it with Ref.~\cite{Geissler_PRB15}. By using Eq. \eqref{eq:velocity} containing the cutoff, we obtain
\begin{equation}\label{eq:Kubo_stat}
\begin{split}
\lim\limits_{\omega\to 0}\langle \hat{I}^{dc}_{BS}\rangle  = {}& 2 \mathrm{e}\gamma(K) \tonde{\frac{\hbar}{E_F}}^{2K} \\  &\tonde{vK -v_F}^2  \left|\omega_0\right|^{2K+1} \sign\left(\omega_0\right) .
\end{split}
\end{equation}
As realized in Ref.~\cite{Geissler_PRB15}, the possibility that $vK \neq v_F$ when interactions are present,  implies a single particle contribution to the backscattering current, which scales as $\sim V^{2K+1}$.
\subsection{Non-interacting limit}
In the non-interacting limit $K=1$, $v=v_F$, the dc component of the current of Eq.~\eqref{eq:Kubo_4} becomes
\begin{alignat}{2}\label{eq:Kubo_K=1}
&&\lim_{K\to 1}\langle \hat{I}^{dc}_{BS}\rangle
=&+\mathrm{e}\frac{\alpha_0^2\,}{2\pi \hbar^2 v^{4 }}\tonde{\omega_0^3 + 3\omega^2\omega_0} + \notag \\
&&& - \mathrm{e}\frac{\alpha_0^2\,}{2\pi \hbar^2 v^{4 }} 2\omega_0 \tonde{\omega^2 + \omega_0^2} +\notag\\
&&& +\mathrm{e}\frac{\alpha_0^2\,}{2\pi \hbar^2 v^{4 }}\omega_0^3  \notag \\
&\Rightarrow\quad
&\lim_{K \to 1}\langle \hat{I}^{dc}_{BS}\rangle
=& \,\frac{\alpha_0^2\,}{2\pi \hbar^3 v_F^{4 }} \omega^2e^2V;.
\end{alignat}
We have used $\Gamma(n+1)=n!$ and taken the limit $a\to 0$. The fermionic calculation in App.~\ref{app:free} confirms the formula above and elucidates its interpretation, which is the following: The backscattering current at second order originates from one-photon resonant processes that transfer electrons from one branch of the linear spectrum to the opposite one, together with a change of energy of $\pm \hbar\omega$. This process is in principle equally probable for both fermionic species. However, for an energy window set by the external bias (of size $eV$) the electrons from the branch with  lower chemical potential energy cannot backscatter into the opposite one because of Pauli principle. Therefore, the net effect is  a decrease of the current originally set by the external bias. This result is in agreement with the results of  Ref.~\cite{Vayrynen_PRL18} for non-interacting QSH edge electrons with a generic spin texture in momentum space. The $\omega^2$ factor in Eq.~\eqref{eq:Kubo_K=1} ensures the absence of backscattering current in the static limit $\omega\to 0$, as   expected~\cite{Geissler_PRB15}.
\section{Conclusions}\label{sec:conclusion}
We have analytically computed the backscattering current due to a $\delta$-like harmonically driven Rashba scatterer in a helical Luttinger liquid, up to second order in the impurity strength. The result is qualitatively different from the case of a time dependent magnetic barrier. It can hence help elucidating the role of impurities on the transport properties of quantum spin Hall systems. Interestingly, our results allow for the definition of an experimentally accessible quantity that only depends on the strength of electron-electron interactions, through the Luttinger parameter $K$. Such a quantity could allow for a measurement of $K$ without the need of measuring power law dependencies.

\begin{acknowledgments}
	This work was supported by the SFB 1170 "ToCoTronics", the W\"{u}rzburg-Dresden Cluster of Excellence ct.qmat (EXC2147, Project No. 39085490), and the Elitenetzwerk Bayern Graduate School on "Topological Insulators". We acknowledge early-stage contributions from S. Barbarino and J.C. Budich.
\end{acknowledgments}

\appendix
\section{Backscattering current operator}\label{app:currop}
In this appendix, we calculate explicitly the commutator in Eq.~\eqref{eq:current_def} that gives the expression for the backscattering current operator, making use of the bosonized version for the density operator $\hat{\rho}_{R} (x) =  \frac{1}{\sqrt{\pi}}\partial_x \Phi_{R}(x)$ and the Rashba Hamiltonian, Eq.~\eqref{eq:Rashba_Ham3}. With the help of  $\quadre{A,e^B} = Ce^B$, if $C=\quadre{A,B}$ and $\quadre{A,C}=\quadre{B,C}=0$, we can compute the first commutator
\begin{equation*}
\quadre{\partial_x \Phi_{R}(x),  e^{i \sqrt{4\pi} \Phi(y)}} =-  \sqrt{\pi}\delta\tonde{x-y} e^{i \sqrt{4\pi} \Phi(y)}\;,
\end{equation*}
where we used standard identities from bosonization~\cite{Miranda_BJP03}. Moreover, we obtain for the other commutator $\quadre{\partial_x \Phi_{R}(x), \partial_y\Theta(y)} =  \frac{i}{2}\partial_y \left( \delta\tonde{x-y}\right) $ directly from the canonical commutation relation of the dual fields. The associated term, when integrated over space, vanishes because we are considering that $\lim_{x\to \pm \infty}\alpha(x) = 0$.
Putting everything together, we  obtain the expression for the backscattering current operator in Eq.~\eqref{eq:curr_op_bos}.
\section{Details on the bosonization calculation}\label{app:bosonization}
We show here the details of the calculation of Eq.~\eqref{eq:curr_pert} in the bosonic language. Upon substituting Eq.~\eqref{eq:Rashba_Ham3} (with $\alpha(x,t) = \alpha_0(t)\delta(x-x_0)$) and~\eqref{eq:curr_op_bos} and using the fact that $\langle \kappa_{L}\kappa_{R}\kappa_{L}\kappa_{R}\rangle = -1$ we obtain
\begin{widetext}
	\begin{equation}\label{eq:Kubo_1}
	\begin{split}
	\langle \hat{I}_{BS}(t)\rangle = \frac{i}{\hbar} \int_{-\infty}^{t} \mathrm{d}t'  & \Biggl\lbrace \sum_{m_1,m_2 = \pm}i \mathrm{e} \frac{2\alpha_0(t)\alpha_0(t')}{\pi \hbar a^2} e^{2i\tonde{m_1+m_2}k_F x_0 }   e^{i\omega_0\tonde{m_1 t'+m_2 t} } \times \\& \times \Biggl[m_2\tonde{\sum_{j=\mathrm{I}}^{\mathrm{IV}}
		C_R^\mathrm{j}(x_0,t;x_0,t'; m_1,m_2)} +\\
	&-\frac{eV}{2\hbar v_F}a\tonde{1 + m_1 m_2}C_R^\mathrm{I}(...) +\frac{e^2V^2}{4\hbar^2 v_F^2} a^2m_1C_R^\mathrm{I}(...) +\\ &-\frac{eV}{2\hbar v_F}a\tonde{C_R^\mathrm{IV}(...) + m_1 m_2C_R^\mathrm{III}(...)}\Biggr] - \rm c.c.\Biggr\rbrace \; .
	\end{split}
	\end{equation}
\end{widetext}
In Eq.~\eqref{eq:Kubo_1}, $(...)$ stands for $(x_0,t;x_0,t'; m_1,m_2)$. The correlation functions are  defined as
\begin{widetext}
	\begin{subequations}\label{eq:c.f.}
		\begin{align}
		C_R^\mathrm{I}(x,t;x',t'; m_1,m_2) = &\frac{ m_1 m_2}{\pi a^2}\bra{0}
		e^{i \sqrt{4\pi}m_1 \Phi(x',t')}
		e^{i m_2\sqrt{4\pi} \Phi(x,t)}  \ket{0} \\	
		C_R^\mathrm{II}(x,t;x',t'; m_1,m_2) = & \;\bra{0}\partial_{x'} \Theta(x',t')
		e^{i m_1\sqrt{4\pi} \Phi(x',t')}\partial_x \Theta(x,t)
		e^{i m_2\sqrt{4\pi} \Phi(x,t)}  \ket{0} \\	
		C_R^\mathrm{III}(x,t;x',t'; m_1,m_2)= & \frac{m_1}{\sqrt{\pi} a}\bra{0}
		e^{i m_1\sqrt{4\pi} \Phi(x',t')}\partial_x \Theta(x,t)
		e^{i m_2\sqrt{4\pi} \Phi(x,t)}  \ket{0}\\
		C_R^\mathrm{IV}(x,t;x',t'; m_1,m_2) = & \frac{m_2}{\sqrt{\pi}a} \bra{0}\partial_x \Theta(x',t')
		e^{i m_1\sqrt{4\pi} \Phi(x',t')}
		e^{i m_2\sqrt{4\pi} \Phi(x,t)}  \ket{0} \;.
		\end{align}
	\end{subequations}
\end{widetext}
Performing standard bosonization calculations~\cite{VonDelft_AOP95,Miranda_BJP03}, we obtain
\begin{widetext}
	\begin{equation*}
	\sum_{j=\mathrm{I}}^{\mathrm{IV}} C_R^\mathrm{j}(x_0,t;x_0,t'; m_1,m_2) = \delta_{m_1,-m_2} \frac{1}{a^2} \frac{2K+1}{2\pi K} \tonde{\frac{t_a}{ t_a +i(t' -t)} }^{2(K+1)} \;,
	\end{equation*}
\end{widetext}
where  $t_a = a/v$ with $v$ the bosonic excitation velocity. 
In general, every correlation function gives a $\delta_{m_1,-m_2}$ factor, so that the fist term in the second line vanishes. We also derive
\begin{subequations}
	\begin{align*}
	&C_R^\mathrm{I}(...) =  \; -\delta_{m_1,-m_2} \frac{1}{\pi a^2} \tonde{\frac{t_a}{ t_a +i(t' -t)} }^{2K} \\	
	&C_R^\mathrm{IV}(...)+ m_1 m_2C_R^\mathrm{III} (...) =  \; -\delta_{m_1,-m_2} \frac{2}{\pi a} \tonde{\frac{t_a}{ t_a +i(t' -t)} }^{2K+1} \;
	\end{align*}
\end{subequations}
and with that
\begin{widetext}
	\begin{equation}\label{eq:Kubo_2}
	\begin{split}
	\langle \hat{I}_{BS}(t)\rangle &=  -\frac{2}{\hbar} \Im \int_{-\infty}^{t} \mathrm{d}t'  i  \mathrm{e}\frac{2\alpha_0(t)\alpha_0(t')}{\pi \hbar a^2}  \Biggl\lbrace \tonde{e^{-i\omega_0\tonde{t'-  t} } - e^{i\omega_0\tonde{t'-  t} }} \times  \\ & \times   \quadre{\frac{2K+1}{2\pi K a^2} \tonde{\frac{t_a}{ t_a +i(t' -t)} }^{2(K+1) } + \frac{e^2V^2}{4\hbar^2 v_F^2} \tonde{\frac{t_a}{ t_a +i(t' -t)} }^{2K}} + \\ &+ \frac{eV}{2\hbar v_F}\tonde{e^{i\omega_0\tonde{t'-  t} } + e^{-i\omega_0\tonde{t'-  t} }}\frac{2}{\pi a} \tonde{\frac{t_a}{ t_a +i(t' -t)} }^{2K+1}   \Biggr\rbrace \; .
	\end{split}
	\end{equation}
\end{widetext}
We now consider a time-periodic impurity $\alpha_0(t)=\alpha_0 \sin(\omega t)$ as in the main text and switch to the integration variable $\tau= t'-t$, in such a way that the dc component of the current is
\begin{widetext}
	\begin{equation}\label{eq:Kubo_3}
	\begin{split}
	\langle \hat{I}^{dc}_{BS}\rangle = {}&  -\mathrm{e}\frac{\alpha_0^2 }{2\pi \hbar^2 a^2} \int_{-\infty}^{\infty}\mathrm{d}\tau    e^{i\omega\tau} \tonde{e^{-i\omega_0\tau } - e^{i\omega_0\tau }} \times \\ & \quad\times \graffe{ \frac{2K+1}{2\pi K a^2}\quadre{ \tonde{\frac{t_a}{ t_a +i\tau} }^{2(K+1)}   - \rm c.c.} + \frac{e^2V^2}{4\hbar^2 v_F^2} \quadre{ \tonde{\frac{t_a}{ t_a +i\tau} }^{2K}   - \rm c.c.}}  + \\
	{}& - \frac{\alpha_0^2 }{\pi^2 \hbar^2 a^3} \frac{eV}{\hbar v_F} \int_{-\infty}^{\infty}\mathrm{d}\tau    e^{i\omega\tau} \tonde{e^{i\omega_0\tau } + e^{-i\omega_0\tau }} \quadre{ \tonde{\frac{t_a}{ t_a +i\tau} }^{2K+1}   + \rm c.c.}\; .
	\end{split}
	\end{equation}
\end{widetext}
If we make use of the integrals
\begin{widetext}
	\begin{subequations}
		\begin{align*}
		&\int_{-\infty}^{\infty}\mathrm{d}x \; e^{i\Omega x}\quadre{\tonde{\frac{t_a}{ t_a +ix} }^{2K+2}   - \tonde{\frac{t_a}{ t_a -ix} }^{2K+2}} =  \frac{2\pi e^{-t_a\left|\Omega\right|} t_a^{2K+2}}{\Gamma(2K+2)} \left|\Omega\right|^{2K+1} \sign\left(\Omega\right)\;, \\	
		&\int_{-\infty}^{\infty}\mathrm{d}x \; e^{i\Omega x}\quadre{\tonde{\frac{t_a}{ t_a +ix} }^{2K}   - \tonde{\frac{t_a}{ t_a -ix} }^{2K}} =  \frac{2\pi e^{-t_a\left|\Omega\right|}t_a^{2K} }{\Gamma(2K)} \left|\Omega\right|^{2K-1} \sign\left(\Omega\right) \;,\\
		&\int_{-\infty}^{\infty}\mathrm{d}x \; e^{i\Omega x}\quadre{\tonde{\frac{t_a}{ t_a +ix} }^{2K+1}   + \tonde{\frac{t_a}{ t_a -ix} }^{2K+1}} =  \frac{2\pi e^{-t_a\left|\Omega\right|} t_a^{2K+1} }{\Gamma(2K+1)} \left|\Omega\right|^{2K} \;,
		\end{align*}
	\end{subequations}
\end{widetext}
defining $t_a= a/v$, we arrive at 
{\begin{widetext}
	\begin{equation}\label{eq:Kubo_4_compl}
	\begin{split}
	\langle \hat{I}^{dc}_{BS}\rangle = {}& \mathrm{e}\frac{\alpha_0^2\,}{2\pi \hbar^2 v^{2(K+1) }a^{2(1-K)} }\frac{2K+1}{ K\,\Gamma(2+2K)} \sum_{r=\pm}re^{-t_a\left|\omega +r\omega_0\right|} \left|\omega + r\omega_0\right|^{2K+1}\sign\left(\omega +r\omega_0\right) + \\
	{}&- \mathrm{e}\frac{\alpha_0^2 }{\pi \hbar^2 v^{2K+1}} \frac{\omega_0}{v_F}\frac{1}{a^{2(1-K)}}\frac{1}{\Gamma(2K+1)} \sum_{r=\pm}e^{-t_a\left|\omega +r\omega_0\right|} \left|\omega +r\omega_0\right|^{2K}  +	 \\
	{}& + \mathrm{e}\frac{\alpha_0^2 }{2\pi \hbar^2 v^{2K}} \frac{\omega_0^2}{2v_F^2} \frac{1}{a^{2(1-K)}}\frac{1}{\Gamma(2K)} \sum_{r=\pm} re^{-t_a\left|\omega +r\omega_0\right|} \left|\omega +r\omega_0\right|^{2K-1}\sign\left(\omega +r\omega_0\right) \;.
	\end{split}
	\end{equation}
\end{widetext}
If we restrict  to a regime where $t_a\left|\omega \pm\omega_0\right| \ll 1$, we obtain Eq.~\eqref{eq:Kubo_4}.}
\section{Free fermion calculation}\label{app:free}
The Rashba Hamiltonian in Eq.~\eqref{eq:Rashba_Ham1} with $\alpha(x,t) = \alpha_0(t)\delta(x-x_0)$ has the momentum space  representation
\begin{equation}\label{eq:Rashba_Ham_k}
\Ham_R = - \frac{\alpha_{0}(t)}{L} \sum_{k_1, k_2} \quadre{i\tonde{k_1+k_2}  \opcdag{k_1 R} \opc{k_2 L } + \rm h.c.} \;.
\end{equation}
Writing the total number of right movers as $\hat{n}_{R} = \sum_{k}\hat{n}_{\mathrm{R} k} = \sum_{k}\opcdag{k R} \opc{k R} $, we can derive the expression for the backscattering current operator again via Eq.~\eqref{eq:current_def}. The result is
\begin{equation}
\hat{I}_{BS} =   -\frac{2\mathrm{e}\alpha_0(t)}{\hbar L}  \sum_{\substack{k_1, k_2 }} (k_1+ k_2)  \tonde{\opcdag{k_1 R}\opc{k_2 L } + \mathrm{h.c.} }   \;.
\end{equation}
The Heisenberg evolution of the fermionic operators according to the clean edge Hamiltonian $\Ham_0 = \hbar v_F\sum_{k} k \tonde{ \hat{n}_R - \hat{n}_L} $ is given by
\begin{equation*}
\begin{cases}
\opc{k R }(t) = e^{-iv_Fkt} \opc{k R} \\
\opc{k L }(t) = e^{iv_Fk  t} \opc{k L } \; .
\end{cases}
\end{equation*}
Using these expressions we can calculate the current with Eq.~\eqref{eq:curr_pert}. By using repeatedly Wick's theorem, we obtain
\begin{widetext}
\begin{equation*}
\begin{aligned}
	\langle \hat{I}_{BS}(t)\rangle
	= &  -\frac{2\mathrm{e}\alpha_0(t)}{L^2\hbar^2}\int_{-\infty}^{0} \mathrm{d}\tau \alpha_0(t+\tau) \sum_{\substack{k_1', k_2}} (k_1+k_2)^2  \tonde{ e^{iv_F(k_1 +k_2 )(\tau)}  \bra {0}\opn{k_1 R } - \opn{k_2 L }\ket{0} +\mathrm{c.c.} } \;,
\end{aligned}
\end{equation*}
\end{widetext}
where $\tau = t'-t$. Writing the impurity matrix element as $\alpha_0(\tau + t)= \alpha_0\sin(\omega t)\cos(\omega \tau) + \alpha_0\cos(\omega t)\sin(\omega \tau)$, we see that only the first term contributes to the dc response. Therefore, after averaging over one driving period and performing the integration over $\tau$, we obtain
\begin{widetext}
\begin{equation}\label{eq:fermions_delta}
\langle \hat{I}_{BS}(t)\rangle =  - \frac{ \mathrm{e}\alpha^2_0}{L^2\hbar^2} \sum_{\substack{k_1', k_2}} (k_1+k_2)^2  \bra {0}\opn{k_1 R } - \opn{k_2 L }\ket{0}  \pi\quadre{\delta\tonde{v_F(k_1+k_2)  +\omega} + \delta\tonde{v_F(k_1+k_2)  -\omega}} \;.
\end{equation}
\end{widetext}
The Dirac delta functions ensure the conservation of energy in the one-photon processes that bring one electron into the opposite branch of the linear spectrum. If we forget about high energy cutoffs, e.g. we consider an infinite linear spectrum, we always get a non-zero result after integrating the Dirac delta functions.
The expectation value with respect to the ground state of the occupation numbers read $\langle\opn{ R k}\rangle = \theta (-k+k^R_{F})$ and
$\langle\opn{ L k}\rangle = \theta (k- k^L_{F})$ respectively. With our choice of chemical potentials, we have $k^R_F =0$ and $k^L_F = - eV/\hbar v_F$. Upon substituting these expressions into the previous equation, we obtain
\begin{equation}
\langle \hat{I}^{dc}_{BS}\rangle = +\frac{ \alpha^2_0\omega^2 e^2 V}{2\pi\hbar^3v_F^4 }  \;,
\end{equation}
which corresponds to Eq.~\eqref{eq:Kubo_K=1} of the main text.
%

\begin{thebibliography}{55}
\bibitem [{Note_magnetic()}]{Note_magnetic}%
\BibitemOpen
\bibinfo {note} {Because a magnetic impurity on the helical edge is
	mathematically equivalent to an ordinary backscatterer in a single-channel
	Luttinger liquid.}\BibitemShut {Stop}%
\end{thebibliography}

\end{document}